\documentclass[amsthm,secthm]{elsart1p}
\usepackage{amsmath, amssymb}

\usepackage{graphicx}
\usepackage{dsfont}
\usepackage{array}
\usepackage[square, comma, sort&compress]{natbib}

\newcommand{\abs}[1]{\lvert #1 \rvert}
\newcommand{\unity}{\mathds{1}}
\newcommand{\Gr}{G^\mathrm{r}}

\begin{document}

\begin{frontmatter}
\title{Optimal block-tridiagonalization of matrices for coherent charge transport}

\author{Michael Wimmer\corauthref{cor}}, \author{Klaus Richter}
\address{Institut f\"ur Theoretische Physik, Universit\"at Regensburg, 
93040 Regensburg, Germany} 
\corauth[cor]{Corresponding author. Email address: 
Michael.Wimmer@physik.uni-regensburg.de}


\begin{abstract}
Numerical quantum transport calculations are commonly based on a tight-binding
formulation. A wide class of quantum transport algorithms requires 
the tight-binding Hamiltonian to be in the form of a block-tridiagonal matrix. 
Here, we develop a matrix reordering algorithm based on graph partitioning 
techniques that yields the optimal block-tridiagonal form for
quantum transport. The reordered Hamiltonian can lead to significant
performance gains in transport calculations, and allows to apply conventional
two-terminal algorithms to arbitrary complex geometries, 
including multi-terminal structures. The block-tridiagonalization algorithm
can thus be the foundation for a generic quantum transport code, applicable to
arbitrary tight-binding systems. We demonstrate the power of this approach by 
applying the block-tridiagonalization algorithm together with the recursive Green's 
function algorithm to various examples of mesoscopic transport in two-dimensional
electron gases in semiconductors and graphene.
 
\end{abstract}

\end{frontmatter}

\section{Introduction}

If the dimensions of a device become smaller than the phase coherence
length $l_\phi$ of charge carriers, classical transport theories 
are not valid any more. Instead, carrier dynamics is now governed by quantum
mechanics, and the wave-like nature of particles becomes important. In general,
the conductance/resistance of such a device does not follow Ohm's law.

In the regime of coherent quantum transport, the Landauer-B\"uttiker
formalism~\citep{Landauer1957,Buttiker1985,Stone1988} relates
the conductance $G$ of a device to the total transmission
probability $T$ of charge carriers through the device,
\begin{equation}
G=\frac{2e^2}{h}\,T=\frac{e^2}{h}\sum_{mn}\abs{t_{mn}}^2\,,
\end{equation} 
where $t_{mn}$ is the transmission amplitude between different
states with transverse quantum numbers $n$ and $m$ in the left and right lead, 
respectively. A state with a given transverse quantum number $n$ 
is also called \emph{channel} $n$.

The problem of calculating the conductance is thus reduced 
to calculating scattering eigenfunctions $\psi$ for a given energy $E$:
\begin{equation}\label{eq:scattering_eigenstate}
(E-H)\psi=0,
\end{equation}
where $H$ is the Hamiltonian of the system. Alternatively, the transmission
probability can be extracted from the retarded Green's function 
$G^\mathrm{r}$ that obeys the equation of motion
\begin{equation}\label{eq:retarded_greenfunction}
(E-H)G^\mathrm{r}=\unity\,.
\end{equation}
The Fisher-Lee relation \citep{Fisher1981,Baranger1989} then allows to calculate the 
transmission ($t_{mn}$) and reflection ($r_{nm}$) amplitudes from $\Gr$. 
In its simplest form, the Fisher-Lee relation reads  
\begin{equation}\label{eq:fisherlee}
t_{mn}=-i \hbar \sqrt{v_{m} v_n}\, \int_{C_\mathrm{R}} dy 
\int_{C_\mathrm{L}} dy'
\phi_{m}(y)\,G^\mathrm{R}(\mathbf{x}, \mathbf{x}')\,\phi_{n}(y')\,,
\end{equation}
and
\begin{equation}\label{eq:fisherlee2}
r_{mn}=\delta_{mn}-i \hbar \sqrt{v_{m} v_n}\, \int_{C_\mathrm{L}} dy 
\int_{C_\mathrm{L}} dy'
\phi_{m}(y)\,G^\mathrm{R}(\mathbf{x}, \mathbf{x}')\,\phi_{n}(y')\,,
\end{equation}
where $v_n$ is the velocity of channel $n$ and the integration runs over the
cross-section $C_\text{L}$ ($C_\text{R}$) of the left (right) lead. 

The Landauer-B\"uttiker formalism can also deal with multi-terminal systems, 
but is restricted to linear response, i.e.~small bias voltages. 
In the general case including external bias, the conductance can be 
calculated using the non-equilibrium Green's function formalism 
(see, e.g.~\citep{Haug1998}).

Except for particularly simple examples, solving 
Eqs.~(\ref{eq:scattering_eigenstate}) and 
(\ref{eq:retarded_greenfunction}) exactly is not possible,
and therefore a numerical computation is often the method of 
choice. Instead of solving directly a differential equation with
its continuous degrees of freedom, such as the Schr\"odinger equation,
numerical computations are usually only attempted within a discrete basis 
set. The differential equation is then replaced by a set of
linear equations, and the Hamiltonian $H$ can be written as a 
matrix. Very often, only few of the matrix elements 
$H_{ij}$ are nonzero. Such \emph{tight-binding}
representations of the Hamiltonian are ubiquitous in quantum
transport calculations and can arise from finite 
differences~\citep{Kimball1934,Pauling1935,Frustaglia2004},
from the finite element method~\citep{Havu2004}, from atomic orbitals
in empirical tight-binding~\citep{Bowen1997,Sanvito1999,Luisier2006}
or Kohn-Sham orbitals within density functional 
theory~\citep{Brandbyge2002,DiCarlo2006,Rocha2006}.

\begin{figure}
\begin{center}
\includegraphics[width=0.8\linewidth]{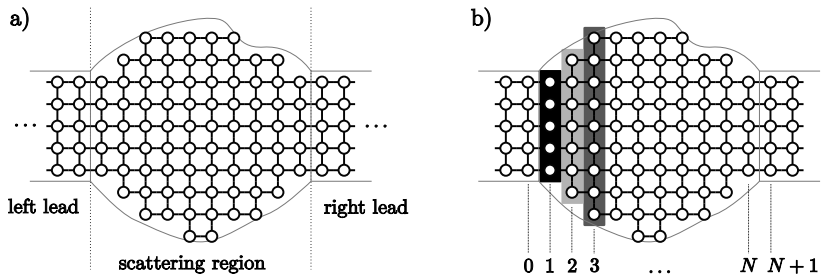}
\end{center}
\caption{(a) Schematic view of a finite difference grid in a two-terminal 
transport setup. (b) Natural ordering of grid points yielding
a block-tridiagonal matrix structure. The different matrix blocks
are marked in alternating shades of grey.}\label{fig:grids}
\end{figure}

When describing transport, the systems under consideration 
are open and thus extend to infinity. 
As a consequence, the tight-binding matrix $H$ is 
infinite-dimensional. However, the conductance 
calculation can be reduced to a finite problem by partitioning the
system into a finite scattering region attached to leads that extend to 
infinity, as schematically depicted in Fig.~\ref{fig:grids}(a). 
For the case of two-terminals, the matrix 
$H$ can be written as
\begin{equation}
H=\left(\begin{array}{ccc}
H_\mathrm{L}&V_\mathrm{LS}&0\\
V_\mathrm{SL}&H_\mathrm{S}&V_\mathrm{SR}\\
0&V_\mathrm{RS}&H_\mathrm{R}
\end{array}
\right)\;,
\end{equation} 
where $H_\mathrm{L(R)}$ is the (infinite) Hamiltonian of the left (right) lead, 
$H_\mathrm{S}$ is the Hamiltonian of the scattering region and of finite size. 
The matrices $V_\mathrm{SL}=V_\mathrm{LS}^\dagger$ and 
$V_\mathrm{SR}=V_\mathrm{rS}^\dagger$ represent the coupling between the 
scattering region and the left and right lead, respectively.

In order to reduce the problem size, it is useful to
introduce the retarded self-energy $\Sigma^\mathrm{r}=\sum_{i=L,R}\; 
V_{\mathrm{S}i}\,g^\mathrm{r}_i\,V_{i\mathrm{S}}$,
where $g^\mathrm{r}_\mathrm{L(R)}$ is the surface Green's function
of the left (right) lead,  i.e.~the value of the Green's function 
at the interface of the lead disconnected from the scattering region.
Then, the Green's function $G_\mathrm{S}$ of the scattering region 
can be calculated as \citep{Datta2002,Ferry2001}
\begin{equation}\label{directinversion}
G^\mathrm{r}_\mathrm{S}=\left(E-H_\text{S}-\Sigma^\mathrm{r}\right)^{-1}\;,
\end{equation}
reminiscent of Eq.~(\ref{eq:retarded_greenfunction}) but with an effective 
Hamiltonian $H_\text{S}+\Sigma^\mathrm{r}$ of finite size.  
This treatment is easily extended to multi-terminal systems. 

Note that it suffices to know the surface Green's function of 
the (semi-)infinite leads, as in a tight-binding Hamiltonian 
the matrices $V_\text{SL}$ and $V_\text{SR}$ have only few nonzero entries. For simple 
systems, the surface Green's function can be calculated analytically
\citep{Datta2002,Ferry2001}, whereas in more complex situations it 
must be computed numerically, either by iteration
\citep{LopezSancho1984,LopezSancho1985}, or by semi-analytical formulas
\citep{Sanvito1999,Krstic2002,Rocha2006}.

The original infinite-dimensional problem has thus been reduced to a finite size matrix 
problem that can, in principle, be solved straight-forwardly on a computer. However, 
for any but rather small problems, the computational task of the direct inversion 
in Eq.~(\ref{directinversion}) is prohibitive. 
Therefore, for two-terminal transport,
many algorithms make use of the \emph{sparsity} of the Hamiltonian matrix 
in tight-binding representation - in particular that this matrix can be written 
in block-tridiagonal form:
\begin{equation}\label{eq:blocktridiagonal}
H=\left(
\begin{array}{@{\extracolsep{4mm}}cccccc@{\extracolsep{0mm}}ccccc}
\ddots&&&\\
&H_\mathrm{L}&V_\mathrm{L}&\\
&V^\dagger_\mathrm{L}&H_\mathrm{L}&H_{0,1}&&&\ddots\\
&&H_{1,0}&H_{1,1}&H_{1,2}&&&0\\
&&&H_{2,1}&H_{2,2}&H_{2,3}&&&\ddots\\
&&&&H_{3,2}&\ddots\\
&&\ddots&&&&\ddots&H_{N-1,N}\\
&&&0&&&H_{N,N-1}&H_{N,N}&H_{N,N+1}\\
&&&&\ddots&&&H_{N+1,N}&H_\mathrm{R}&V_\mathrm{R}\\
&&&&&&&&V^\dagger_\mathrm{R}&H_\mathrm{R}\\
&&&&&&&&&&\ddots
\end{array}\right)\;,
\end{equation}
where the index $\mathrm{L}$ ($\mathrm{R}$) denotes the blocks in the left (right) lead,
$1 \dots N$ the blocks within the scattering region, and $0$ ($N+1$) the first block in
the left (right) lead. Such a form arises, for example,
naturally in the method of finite differences, 
when grid points are grouped into vertical slices according to their $x$-coordinates, 
as shown in Fig.~\ref{fig:grids}(b), but also applies to any other sparse tight-binding 
Hamiltonian. 

The block-tridiagonal form of the Hamiltonian is the foundation
of several quantum transport algorithms for two-terminal systems.
The transfer matrix approach applies naturally to block-tridiagonal
Hamiltonians, but becomes unstable for larger systems. However, a stabilized version
has been developed by Usuki \emph{et al.}~\citep{Usuki1994,Usuki1995}. 
In the decimation technique~\citep{Lambert1980,Leadbeater1998},
the Hamiltonian of the scattering region is replaced by an effective Hamiltonian
between the two leads by eliminating internal degrees of freedom. The contact block 
reduction method~\citep{Mamaluy2005} calculates the full Green's function of the system 
using a limited set of eigenstates. The recursive Green's function (RGF) 
technique~\citep{Thouless1981,Lee1981,MacKinnon1985}
uses Dyson's equation to build up the system's Green's function block by block. 
It has also been adapted to Hall geometries with four terminals~\citep{Baranger1991} and
to calculate non-equilibrium densities~\citep{Lake1997,Lassl2007}. Furthermore, the 
RGF algorithm has been formulated to be suitable for parallel 
computing~\citep{Drouvelis2006}.

\begin{figure}
\center
\includegraphics[width=0.8\linewidth]{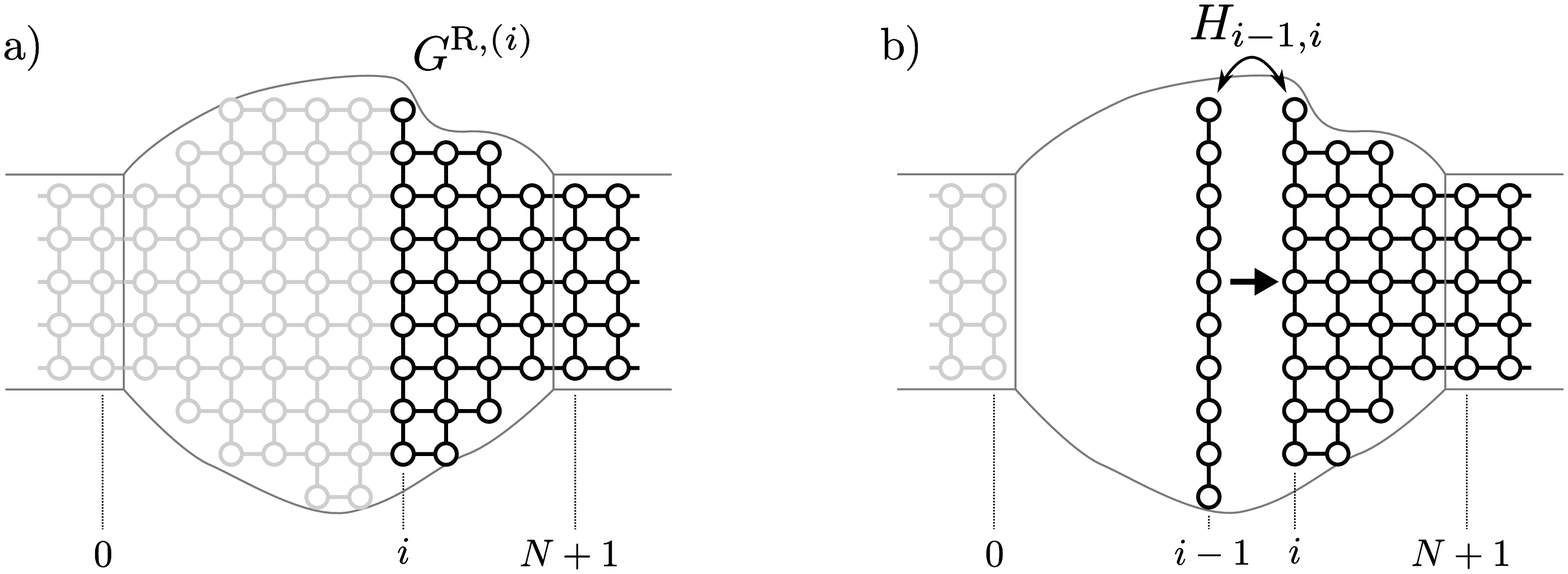}
\caption{Schematic depiction of the recursive Green's function algorithm:
(a) The Green's function $G^{\mathrm{r},(i)}$ contains all blocks
$\geq i$. (b)  The Green's function $G^{\mathrm{r},(i-1)}$ is obtained
by adding another matrix block.}\label{fig:rgf}
\end{figure}

Of course, there are also other transport techniques not
directly based on the block-tridiagonal form of the Hamiltonian matrix, such as 
extracting the Green's function from wave packet dynamics \citep{Kramer2008}. Still,
such algorithms are not as widely used as the large class of algorithms, that are
directly based on the block-tridiagonal form of the Hamiltonian.
In order to illustrate the typical computational tasks of this class of algorithms, 
we briefly explain, as a representative example, the RGF algorithm. 

The RGF technique is based on Dyson's equation $\Gr=\Gr_0+\Gr_0 V \Gr$ 
(see, e.g~\citep{MacKinnon1985}), 
where $\Gr$ denotes the Green's function of the perturbed system, $\Gr_0$ that of the 
unperturbed system and $V$ the perturbation. Using this equation, the
system is built up block by block, as depicted in Fig.~\ref{fig:rgf}. Let
$G^{\mathrm{r},(i)}$ denote the Green's function for the system containing 
all blocks $\geq i$. Then, at energy $E$, the Green's function 
$G^{\mathrm{r},(i-1)}$ is related to $G^{\mathrm{r},(i)}$ by
\begin{equation} 
G^{\mathrm{r},(i-1)}_{i-1,i-1}=\left(E-H_{i-1,i-1}-
H_{i-1,i}\;G^{\mathrm{r},(i)}_{i,i}\;H_{i,i-1}\right)^{-1}
\end{equation}
and
\begin{equation} 
G^{\mathrm{r},(i-1)}_{N+1,i-1}=G^{\mathrm{r},(i)}_{N+1,i}\;H_{i,i-1}\;G^{\mathrm{r},(i-1)}_{i-1,i-1}\;.
\end{equation}
Starting from $G^{\mathrm{r},(N+1)}_{N+1,N+1}=g^{\mathrm{r}}_\mathrm{R}$, 
the surface Green's function of the right lead, $N$ slices are added 
recursively, until $G^{\mathrm{r},(1)}$ has been calculated. The blocks 
of the Green's function of the full system necessary for transport are 
then given by
\begin{equation}  
G^{\mathrm{r}}_{0,0}=\left(\left(g^\mathrm{r}_\mathrm{L}\right)^{-1}-\;H_{0,1}\;G^{\mathrm{r},(1)}_{1,1}\;H_{1,0}\right)^{-1}
\end{equation}
and
\begin{equation}
G^{\mathrm{r}}_{N+1,0}=G^{\mathrm{r},(1)}_{N+1,1}\;H_{1,0}\;G^{\mathrm{r}}_{0,0}\;,
\end{equation}
where $g_\mathrm{L}^\mathrm{r}$ is the surface Green's function of the left lead.
$G^{\mathrm{r}}_{0,0}$ and $G^{\mathrm{r}}_{N+1,0}$ are sufficient to calculate 
transmission and reflection probabilities via the Fisher-Lee 
relation, Eqs.~\eqref{eq:fisherlee} and \eqref{eq:fisherlee2}.

Each step of the algorithm performs inversions and matrix multiplications 
with matrices of size $M_i$. Since the computational complexity of 
matrix inversion and multiplications scales as $M_i^3$, the complexity of 
the RGF algorithm is $\propto \sum_{i=0}^{N+1} M_i^3$. Thus, it
scales linearly with the ``length'' N, and cubically with the 
``width'' $M_i$ of the system. This scaling also applies to most of the 
other transport algorithms  mentioned above.

While for particular cases general transport algorithms, such as the
RGF algorithm, cannot compete with more 
specialized algorithms, such as the modular recursive Green's function 
technique~\citep{Rotter2000,Rotter2003} that is optimized for
special geometries, they are very versatile and easily adapted to 
two-terminal geometries---provided that the leads are arranged collinearly. 
Amongst other things, this restriction will be lifted by the 
approach presented in this work.

Although the block-tridiagonal structure of $H$, 
Eq.~(\ref{eq:blocktridiagonal}), that arises naturally in many problems 
appears to have a small ``width'' and thus seems to be quite suitable for 
transport algorithms, optimizing the block-tridiagonal structure by 
reordering the matrix $H$ may lead to a significant
speed-up in the conventional two-terminal algorithms, as we show below. 
Furthermore, such a reordering allows for the application of the 
established two-terminal algorithms to more complex geometries, such
as non-collinear leads or multi-terminal structures, that would otherwise
need the development of specialized algorithms.

Below, we develop a matrix reordering algorithm based on graph 
partitioning techniques that brings an arbitrary matrix $H$ into
a block-tridiagonal form optimized for transport calculations. To this end, 
the paper is organized as follows. In section~\ref{section:algorithm}
we formulate the matrix reordering problem in the language of graph
theory and develop the reordering algorithm. In section \ref{section:examples}
we apply this algorithm to various examples and investigate its performance and
the performance of the RGF algorithm for the reordered Hamiltonian $H$.
We conclude in section~\ref{section:conclusions}.

\section{Optimal Block-tridiagonalization of matrices}\label{section:algorithm}

\subsection{Definition of the problem}

\subsubsection{Definition of the matrix reordering problem}

As shown in the introduction, the typical runtime of transport algorithms,
is proportional to $\sum_{i=0}^{N+1} M_i^3$, does depend on the particular block-tridiagonal
structure of $H$. Therefore, the runtime of these algorithms can be improved
in principle by conveniently reordering $H$ with a permutation $P$,
\begin{equation}
H'=P\;H\;P^{-1}\;.
\end{equation}

In order to quantify how a typical transport algorithm performs
for a given matrix structure,  we define a weight $w(H)$ associated 
with a matrix $H$ as 
\begin{equation}\label{eq:matrix_weights}
w(H)=\sum_{i=0}^{N+1} M_i^3\,,
\end{equation}
where $M_i$ is the size of block $H_{i,i}$.
Optimizing the matrix for transport algorithms is then equivalent to minimizing the 
weight $w(H)$. Since $\sum_{i=0}^{N+1} M_i = N_\mathrm{grid}$, where $N_\mathrm{grid}$ 
is the total number of grid points, $w(H)$ is minimal, if all $M_i$ are equal, 
$M_i=N_\mathrm{grid}/(N+2)$. Therefore, a matrix tends to have small weight, if the number $N$ of
blocks is large, and all blocks are equally sized. The reordering problem
of the matrix $H$ is thus summarized as follows:

\begin{prob}\label{matrix_reordering_problem}
{\bf Matrix reordering problem:}
Find a reordered matrix $H'$ such, that
\begin{enumerate}
\item $H'_{00}$ and $H'_{N+1N+1}$ are blocks given by the left and
right leads (as required by transport algorithms)
\item $H'$ is block-tridiagonal ($H'_{ij}\neq 0$, iff $j=i+1,i,i-1$),
\item the number $N$ of blocks is as large as possible, and
all blocks are equally sized.
\end{enumerate}
\end{prob}

In principle, this constrained optimization problem could be solved 
by generic optimization algorithms, such as \emph{Simulated Annealing}.
However, for larger problems the optimization could take much more time
than the actual transport calculation, rendering the optimization process
useless. It is therefore necessary to use heuristics especially
designed for the problem at hand. To this end, we formulate the matrix 
reordering problem in the language of graph theory.

\subsubsection{Mapping onto a graph partitioning problem}

A \emph{graph} $\mathcal{G}$ is an ordered pair 
$\mathcal{G}=(\mathcal{V}, \mathcal{E})$, where $\mathcal{V}$ is a set of 
\emph{vertices} $v$ and  $\mathcal{E}$ a set of ordered pairs of vertices $(v_1,v_2) 
\in \mathcal{V}\times \mathcal{V}$. Such a pair is called an \emph{edge}. A graph is 
called \emph{undirected}, if for every edge $(v_1,v_2)\in \mathcal{E}$ 
also $(v_2,v_1)\in \mathcal{E}$. Two vertices $v_1$ and $v_2$  are called \emph{adjacent}, if
$(v_1,v_2)\in \mathcal{E}$. In order to simplify the notation, we will also
consider a vertex $v$ to be adjacent to itself. 

There is a natural one-to-one correspondence between graphs and the structure
of sparse matrices. For a given $n\times n$ matrix $H$, we define a graph
$\mathcal{G}=(\mathcal{V},\mathcal{E})$ with 
$\mathcal{V}=\{1,\dots, n\}$ and $(i,j)\in \mathcal{E}$ iff the entry $h_{ij} \neq 0$.
A graph thus stores information about the \emph{structure} of a matrix, 
i.e.~which entries are nonzero. It does not contain any information about the values
of the respective entries, although these may be stored easily along with the
graph. However, for the formulation of the quantum transport algorithms,
only the block-tridiagonal form, i.e.~the structure of the matrix, is relevant.  
Hermitian matrices, that are considered in quantum transport, have a symmetric
structure of zero and nonzero entries, and therefore the corresponding graphs are
undirected.

\begin{figure}
\center
\includegraphics[width=0.9\linewidth]{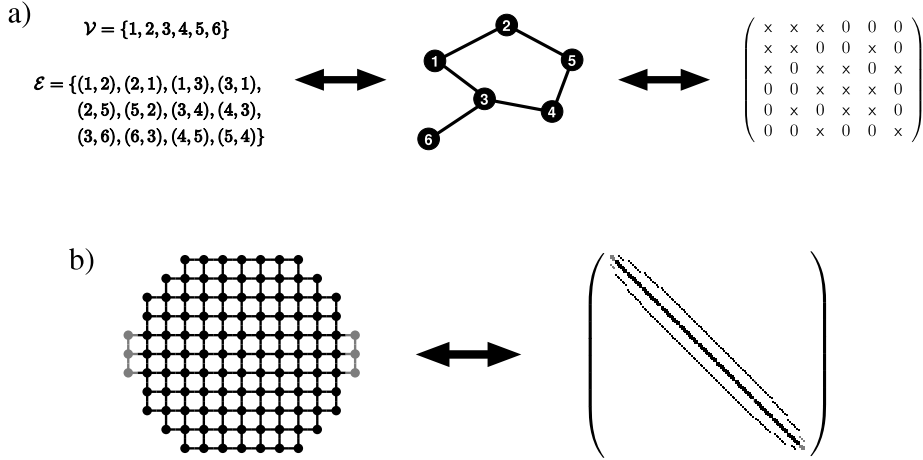}
\caption{(a) Simple example showing the connection between a graph, its 
graphical representations with dots and lines, and the zero--nonzero structure
of a matrix. (b) Example of a finite difference grid, that can be interpreted as
a graph, and the structure of the corresponding matrix. Nonzero entries are shown
as black dots.}\label{fig:graphs}
\end{figure}

A graph can be depicted by drawing
dots for each vertex $v$, and lines connecting these dots for every 
edge $(v_1,v_2)$, as shown in Fig.~\ref{fig:graphs}(a). It should be noted 
that a graphical representation of a tight-binding grid, such as shown in
Fig.~\ref{fig:graphs}(b), can be directly interpreted as a 
representation of a graph and the corresponding matrix structure. 

In terms of graph theory, 
matrix reordering corresponds to renumbering the vertices of a graph.
Since we are only interested in reordering the matrix
in terms of matrix blocks (the order within a block should not
matter too much), we define a \emph{partitioning} of $\mathcal{G}$
as a set $\{\mathcal{V}_i\}$ of disjoint subsets $\mathcal{V}_i \subset \mathcal{V}$ such that
$\bigcup_{i} \mathcal{V}_i = \mathcal{V}$ and $\mathcal{V}_i \cap \mathcal{V}_j =\emptyset$ for $i\neq j$. 
Using these concepts, we can now reformulate the original matrix reordering
problem into a graph partitioning problem:

\begin{prob}\label{graph_part_problem}
{\bf Graph partitioning problem:} Find a partitioning $\{\mathcal{V}_0, \dots, \mathcal{V}_{N+1}\}$
of $\mathcal{G}$ such that:
\begin{enumerate}
\item\label{req1} $\mathcal{V}_0$ and $\mathcal{V}_{N+1}$ contain the vertices belonging to left and right leads,
\item\label{req2}
\begin{enumerate}
\item vertices in $\mathcal{V}_0$ and $\mathcal{V}_{N+1}$ are only connected to 
vertices in $\mathcal{V}_1$ and $\mathcal{V}_N$, respectively,
\item  for $0<i<N+1$, there are edges between $\mathcal{V}_i$ and $\mathcal{V}_j$ iff $j=i+1,i,i-1$,
\end{enumerate}
\item\label{req3} the number $N+2$ of sets $\mathcal{V}_i$ is as large as possible, and all sets $\mathcal{V}_i$ have the
same cardinality $\left|\mathcal{V}_i\right|$. A partitioning with all $\left|\mathcal{V}_i\right|$ equally sized is called \emph{balanced}. 
\end{enumerate} 
\end{prob}

A partitioning obeying requirement \ref{graph_part_problem}.\ref{req2} 
is called a \emph{level set} with \emph{levels} $\mathcal{V}_i$~\citep{Gibbs1976}. Level sets appear
commonly as an intermediate step in algorithms for bandwidth reduction of
matrices~\citep{Gibbs1976,Cuthill1969,George1971,Liu1976}. These algorithms seek
to find a level set of minimal width, i.~e. $\max_{i=0\dots N+1} |\mathcal{V}_i|$ as small as possible
which is equivalent to requirement \ref{graph_part_problem}.\ref{req3}. 
The main difference between our graph 
partitioning problem and the bandwidth reduction problem is requirement 
\ref{graph_part_problem}.\ref{req1}: In
the graph partitioning problem, $\mathcal{V}_0$ and $\mathcal{V}_N$ are determined by the problem at hand, 
while in the bandwidth reduction problem these can be chosen freely.
Due to this difference, bandwidth reduction algorithms can be applied successfully
to our graph partitioning problem only for special cases, as we show below.

The term \emph{graph partitioning} usually refers to the general problem of finding a balanced 
partitioning $\{\mathcal{V}_i\}$ of a graph and has many  applications
in various fields such as very-large-scale integration (VLSI) 
design \citep{Kernighan1970,Fiduccia1982,graph:Karypis00}, 
sparse matrix reorderings for LU or Cholesky decompositions \citep{Hendrickson1998}, or  
block ordering of sparse matrices for parallel computation 
\citep{O'Neil1990,Coon1995,Camarda1999,Hendrickson1998a,Hendrickson2000,Aykanat2004}.
In particular, the latter examples also include block-tridiagonal 
orderings \citep{Coon1995,Camarda1999}. However, as these reorderings are geared towards
parallel computation, they obtain a fixed number $N$ of sets $\mathcal{V}_i$ given by the number 
of processors of a parallel computer, whereas in our block-tridiagonal reordering 
the number $N$ should be as large as possible. In addition to that, the constraints on
the blocks $\mathcal{V}_0$ and $\mathcal{V}_{N+1}$ (requirement \ref{graph_part_problem}.\ref{req1}) are again not present there.

As we cannot directly employ existing techniques to solve the graph partitioning problem,
we will develop an algorithm combining ideas from both bandwidth reduction and graph 
partitioning techniques in the subsequent sections: Concepts from bandwidth reduction 
are employed to construct a level set which is then balanced using concepts from graph 
partitioning.

\subsection{Matrix reordering by graph partitioning}

\subsubsection{A local approach---breadth first search}

A breadth-first-search (BFS) \citep{Sedgewick1992} on a graph immediately yields a level set
\citep{Gibbs1976,Cuthill1969,George1971,Liu1976}. In our particular example, the level set
is constructed as follows:

\begin{alg}\label{BFSalgo}
Level set construction by breadth-first-search.

\begin{enumerate}
\item[A] Start from $i=0$. Then, $\mathcal{V}_i=\mathcal{V}_0$, as the first level is given by the constraints
of requirement (\ref{req1}).
\item[B] If there is a vertex in $\mathcal{V}_i$ that is adjacent to a vertex in $\mathcal{V}_{N+1}$, 
assign all the remaining unassigned vertices into $\mathcal{V}_i$ and end the algorithm. 
\item[C] All vertices adjacent to $\mathcal{V}_i$ that are not contained in the previous levels 
$\mathcal{V}_i, \mathcal{V}_{i-1},\dots \mathcal{V}_0$ are assigned to $\mathcal{V}_{i+1}$.  
\item[D] Continue at step B with $i=i+1$.
\end{enumerate}
\end{alg}

Note that the sets $\{\mathcal{V}_i\}$ form a level set by construction---a set $\mathcal{V}_i$ may only
have vertices adjacent to $\mathcal{V}_{i-1}$ and $\mathcal{V}_{i+1}$. The construction by BFS 
not only obtains the number of levels $N+2$ for a particular realization, but yields
a more general information: 
\begin{lem}
The number of levels $N+2$ in the level set constructed by algorithm \ref{BFSalgo}
is the maximum number of levels compatible with the constraints on the initial
and final level $\mathcal{V}_0$ and $\mathcal{V}_{N+1}$ for a graph $\mathcal{G}$.
\end{lem}
This can
be seen from the fact that a BFS finds the shortest path in the graph between the initial
sets $\mathcal{V}_0$ and $\mathcal{V}_{N+1}$, $(v_0, v_1, \dots, v_i, \dots, v_{N+1})$ where $v_0 \in \mathcal{V}_0$
and $v_{N+1} \in \mathcal{V}_{N+1}$. Any vertex on this shortest path can be uniquely assigned to
a single level $\mathcal{V}_i$ and it would not be compatible with a larger number of levels 
than $N+2$. 

Algorithm \ref{BFSalgo} not only yields the maximum number of levels: All vertices
contained in the first $n$ levels of the BFS must be contained in the first $n$ 
levels of \emph{any} other level set.
\begin{lem}\label{lemma_minimal_set}
Let $\{\mathcal{V}_0,\mathcal{V}_1,\dots,\mathcal{V}_{N+1}\}$ be a level set constructed by algorithm \ref{BFSalgo}, and
$\{\mathcal{V}'_0, \mathcal{V}'_1, \dots, \mathcal{V}'_{N'+1}\}$ another level set consistent with the requirements of
problem \ref{graph_part_problem} with $N'\leq N$. 
Then $\mathcal{V}_0\cup \mathcal{V}_1\cup\dots\cup \mathcal{V}_n \subset 
\mathcal{V}'_0\cup \mathcal{V}'_1\cup\dots\cup \mathcal{V}'_n$ for $0\leq n\leq N'+1$.
\end{lem}
The statement is proved by induction. It is true trivially for $n=0$ (because
of requirement \ref{req1} in problem \ref{graph_part_problem}) and for
$n=N'+1$ (then the levels cover the whole graph). Suppose now that the statement 
holds for $n<N'$. Note that for the proof it suffices to show that 
$\mathcal{V}_{n+1} \subset \mathcal{V}'_0\cup \mathcal{V}'_1\cup\dots,\cup \mathcal{V}'_{n+1}$.
Consider now the set of all vertices adjacent to 
$\mathcal{V}_n$, $\text{adjacent}(\mathcal{V}_n)=
\{v \in \mathcal{V}\; |\; v \text{ is adjacent to some } v' \in \mathcal{V}_n\}$.
By construction, $\mathcal{V}_{n+1} \subset \text{adjacent}(\mathcal{V}_n)$. Since 
$\mathcal{V}_n \subset \mathcal{V}'_0\cup \mathcal{V}'_1\cup\dots\cup \mathcal{V}'_n$ 
and ${\mathcal{V}'_i}$ is a level set, all
vertices adjacent to $\mathcal{V}_n$ must be contained in the set of
vertices including the next level, i.e.~$\text{adjacent}(\mathcal{V}_n) \subset 
\mathcal{V}'_0\cup \mathcal{V}'_1\cup\dots,\cup \mathcal{V}'_n\cup \mathcal{V}'_{n+1}$. But then also 
$\mathcal{V}_{n+1} \subset \mathcal{V}'_0\cup \mathcal{V}'_1\cup\dots,\cup \mathcal{V}'_{n+1}$, 
which concludes the proof.

\begin{figure}
\begin{center}
\includegraphics[width=0.42\linewidth]{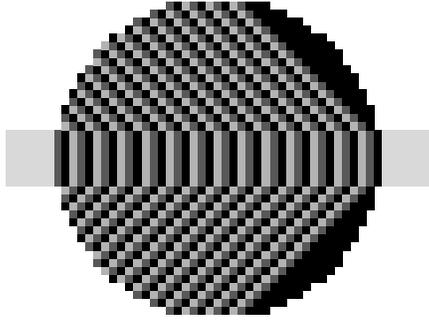}
\end{center}
\caption{Level set created by a BFS starting from $\mathcal{V}_0$. Different levels are 
shown in alternating shades of grey.}\label{example_BFS}
\end{figure}

Thus, the vertices contained in the first $n$ levels of the BFS form a minimal set
of vertices needed to construct $n$ levels. However, this also implies that the last level
which then covers the remaining vertices of the graph, may contain
many more vertices than the average, leading to an unbalanced level set. 
This is not surprising, since the algorithm does not explicitly consider
balancing and only local information is used, i.e.~whether a vertex is adjacent 
to a level or not. An example for this imbalance is shown in Fig.~\ref{example_BFS}, 
the BFS construction yields a very large last level. 

Note that throughout the manuscript we visualize the graph theoretical 
concepts using examples of graphs obtained from discretizing a two-dimensional structure. 
However, the ideas and algorithms presented here apply to any graph
and are not limited to graphs with coordinate information. Two-dimensional
graphs have the advantage of being visualized easily. In particular,
the BFS search has an intuitive physical analog: Wave front
propagation of elementary waves emanating from the vertices of
the initial level $\mathcal{V}_0$.

The problem that a BFS does not yield a balanced partitioning was also noted
in the theory of bandwidth reduction. The Gibbs-Poole-Stockmeyer (GPS) algorithm
tries to overcome this deficiency by constructing a level set
through the combination of two BFS searches starting from the
initial and the final levels. However there the initial and final levels 
are sought to be furthest apart, contrary to our problem. In general,
the GPS construction only yields a balanced level set 
if the initial and final level are close to furthest apart, as we will show in 
Sec.~\ref{section:examples}.

\subsubsection{A global approach---recursive bisection}

In order to obtain a balanced partitioning, graph partitioning algorithms
commonly perform a recursive bisection, i.e.~successively bisect the graph
and the resulting parts until the desired number of parts is obtained
\citep{Kernighan1970,Fiduccia1982,Coon1995,Camarda1999,Gupta1997,Simon1997}. 
This approach has the advantage of reducing the partitioning problem to 
a simpler one, i.e.~bisection. Furthermore, if the resulting parts of every
bisection are equally sized, the overall partitioning will be balanced. 
In addition, bisection is inherently a global approach, as the whole
graph must be considered for splitting the system into two equally sized parts.
Thus, it can be expected
to yield better results than a local approach, such as BFS.

\begin{figure}
\includegraphics[width=\linewidth]{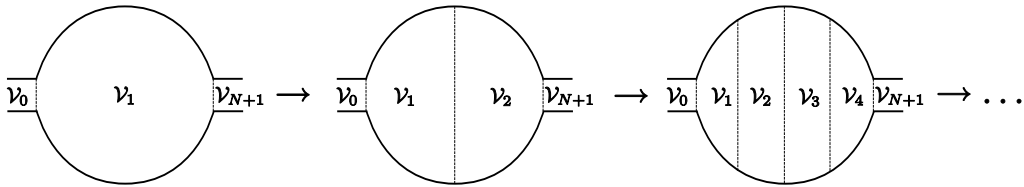}
\caption{Schematic depiction of recursive bisection.}\label{fig:recursive_bisection_example}
\end{figure}

We intent to construct a level set with $N+2$ levels, where $N+2$ is the maximum
number of levels as determined by algorithm \ref{BFSalgo}. To this end we start from an
initial partitioning $\{\mathcal{V}_0,\mathcal{V}_1,\mathcal{V}_{N+1}\}$, where $\mathcal{V}_0$ and $\mathcal{V}_{N+1}$ contain the vertices
of the leads (requirement \ref{graph_part_problem}.\ref{req1}), and $\mathcal{V}_1$ all other vertices.
The level set is then obtained by applying the bisection algorithm recursively
to $\mathcal{V}_1$ and the resulting subsets, until $N$ levels are obtained, as shown
schematically in Fig.~\ref{fig:recursive_bisection_example}.
Here bisection means splitting a set $\mathcal{V}_i$ into two sets, $\mathcal{V}_{i_1}$ and $\mathcal{V}_{i_2}$,
such that $\mathcal{V}_{i_1}\cup \mathcal{V}_{i_2} = \mathcal{V}_i$ and $\mathcal{V}_{i_1}\cap \mathcal{V}_{i_2} = \emptyset$.
In oder to be applicable to the graph partitioning problem \ref{graph_part_problem},
the bisection must comply with certain requirements: 
\begin{prob}\label{bisection_prob}
The bisection algorithm must be
\begin{enumerate}
\item\label{bisect_req1} compatible with a level set with $N+2$ levels.
\item\label{bisect_req2} balanced.
\item\label{bisect_req3} performed such that subsequent bisections may 
lead to a balanced level set.
\end{enumerate}
\end{prob}
Requirement \ref{bisection_prob}.\ref{bisect_req3} is formulated rather vaguely: Usually
there are many different choices how to perform a bisection. A particular choice will
influence the subsequent bisections (for a similar problem in graph partitioning
see \citep{Simon1997}), and thus the bisection algorithm must in principle 
take all following bisection steps into account. Since an exact solution to that 
problem seems computationally intractable, we will resort to heuristics there.

We start explaining how to comply with requirements 
\ref{bisection_prob}.\ref{bisect_req1} and \ref{bisection_prob}.\ref{bisect_req2}.
In the following we assume that $N>0$, as $N=-1,0$ are trivial cases. Then
the initial partitioning $\{\mathcal{V}_0,\mathcal{V}_1,\mathcal{V}_{N+1}\}$ forms a level set, and so will the final
result of the recursive bisection, if the result of every intermediate
bisection yields a level set. For this, consider a set $\mathcal{V}_i$ with vertices adjacent to the sets
$\mathcal{V}_{i_\text{left}}$ and $\mathcal{V}_{i_\text{right}}$, where ``left''(``right'') is defined
as being closer to $\mathcal{V}_0$ ($\mathcal{V}_{N+1}$). Then the sets resulting from the bisection,
$\mathcal{V}_{i_1}$ and $\mathcal{V}_{i_2}$ may only have vertices adjacent to $\mathcal{V}_{i_\text{left}}$,$\mathcal{V}_{i_2}$
and $\mathcal{V}_{i_1}$,$\mathcal{V}_{i_\text{right}}$, respectively.

Apart from the condition of forming a level set, requirement 
\ref{bisection_prob}.\ref{bisect_req1} also dictates the total number of
levels. Due to the nature of the recursive bisection, the number of final levels contained
in an intermediate step is always well-defined. If a set $\mathcal{V}_i$ contains $N_i$
levels, then $\mathcal{V}_{i_1}$ and $\mathcal{V}_{i_2}$ must contain
$N_{i_1}=\text{Int}(N_i/2)$ and $N_{i_2}=N_i-\text{Int}(N_i/2)$ levels, respectively. Here,
$\text{Int}(\dots)$ denotes rounding off to the next smallest integer. The bisection
is thus balanced, if 
\begin{equation}\label{eq:balance_criterion}
\abs{\mathcal{V}_{i_1}} \approx \frac{N_{i_1}}{N_i}\; \abs{\mathcal{V}_i} \quad\text{and}\quad 
\abs{\mathcal{V}_{i_2}}\approx \frac{N_{i_2}}{N_i}\; \abs{\mathcal{V}_i}\,.
\end{equation}
Note that $N_i$ can take any value, and usually is not a power of two. 

\begin{figure}
\includegraphics[width=\linewidth]{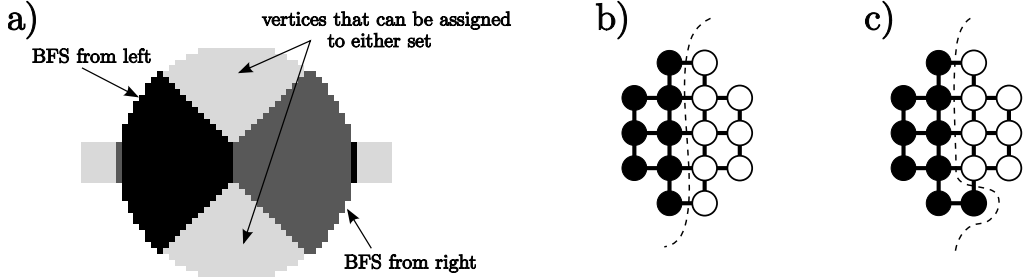}
\caption{(a) Example showing for a disk-type geometry the BFS from the left and
right neighboring sets that construct the minimal set of vertices
$\mathcal{V}_{i_1,\text{BFS}}$ (black) and $\mathcal{V}_{i_2,\text{BFS}}$ (dark grey) that must be 
contained in $\mathcal{V}_{i_1}$ and $\mathcal{V}_{i_2}$, respectively. The remaining vertices (light grey)
can be assigned to either set. (b) and (c): Examples illustrating the difference between
cut edges and cut nets. The number of cut edges is 5 in both (b) and (c),
while the number of cut nets (boundary vertices) is 10 in (b) and 9 in (c).
}\label{fig:bisection_explanation}
\end{figure}

From Lemma \ref{lemma_minimal_set} we know that the minimum set of vertices necessary
to form $n$ levels is given by a BFS up to level $n$. Let $\mathcal{V}_{i_1,\text{BFS}}$ 
($\mathcal{V}_{i_2,\text{BFS}}$) denote the set of vertices found by a BFS starting
from $\mathcal{V}_{i_\text{left}}$ ($\mathcal{V}_{i_\text{right}}$) up to level $N_{i_1}$ ($N_{i_2}$).
Then, for any bisection complying with requirement \ref{bisection_prob}.\ref{bisect_req1},
$\mathcal{V}_{i_1,\text{BFS}} \subset \mathcal{V}_{i_1}$ and $\mathcal{V}_{i_2,\text{BFS}} \subset \mathcal{V}_{i_2}$. These
vertices are uniquely assigned to $\mathcal{V}_{i_1}$ and $\mathcal{V}_{i_2}$ and are consequently
marked as \emph{locked}, i.e.~later operations may not change this
assignment. An example for the vertices found in a BFS is shown in 
Fig.~\ref{fig:bisection_explanation}(a). Note that
in the initial bisection, $\mathcal{V}_i=\mathcal{V}_1$, $N_i=N$, $\mathcal{V}_{i_\text{left}}=\mathcal{V}_0$,
and $\mathcal{V}_{i_\text{left}}=\mathcal{V}_{N+1}$. 

The remaining unassigned vertices can be assigned to 
either set, and the bisection will still be compatible
with a level set containing $N+2$ vertices. Thus for complying with requirement
\ref{bisection_prob}.\ref{bisect_req2}, any prescription obeying the balancing criterion
may be used. We choose to distribute the remaining vertices by continuing
the BFS from $\mathcal{V}_{i_\text{left}}$ and $\mathcal{V}_{i_\text{right}}$ and assigning vertices to
$\mathcal{V}_{i_1}$ and $\mathcal{V}_{i_2}$ depending on their distance to the left or right
neighboring set, while additionally obeying the balancing criterion. This 
approach---assigning vertices to levels according to their distance from the initial
and final set---is rather intuitive and probably the procedure that would be
used if the level set were to be constructed ``by hand''. This procedure may lead
to reasonable level sets, however in general, additional optimization on the 
sets $\mathcal{V}_{i_1}$ and $\mathcal{V}_{i_2}$ is needed, as discussed below. If this optimization is
used, it can also be useful to distribute the unassigned vertices randomly, as this may
help avoiding local minima.

As mentioned above, there is a lot of arbitrariness in distributing the unassigned
vertices into $\mathcal{V}_{i_1}$ and $\mathcal{V}_{i_2}$. However, the particular choice of the bisection
will influence whether a later bisection is balanced or not: If $\mathcal{V}_{i_{1}(i_2),\text{BFS}}$
contains more vertices than the balance criterion \eqref{eq:balance_criterion},
the bisection \emph{cannot} be balanced. Obviously, the BFS that constructs
$\mathcal{V}_{i_{1}(i_2),\text{BFS}}$ depends on the details of the set $\mathcal{V}_i$ and thus
on the details of the previous bisection step.

In order to formulate a criterion that may resolve the above mentioned arbitrariness
and help to find a balanced level set, it is useful to consider the matrix representation
of the graph $\mathcal{G}$. Bisecting a graph means ordering the corresponding matrix into two
blocks that are connected by an off-diagonal matrix $H_{i_1,i_2}$:
\begin{equation}
\left(
\begin{array}{ccc|ccc}
&\ddots&&\\
&&&H_{i_1,i_2}&\\
\hline
&&H_{i_2,i_1}&&\\
&&&&\ddots\\
\end{array}
\right)\,.
\end{equation}
This off-diagonal matrix will be unchanged by further bisections and thus determines
the minimum level width that can be achieved. Therefore, the size of the 
off-diagonal matrix $H_{i_1,i_2}$ should be minimized. 

In a bisection, an edge $(v_1,v_2) \in \mathcal{E}$ is said to be \emph{cut}, if
$v_1$ and $v_2$ belong to different sets, i.e.~$v_1 \in \mathcal{V}_{i_1}$ and $v_2 \in \mathcal{V}_{i_2}$
or vice versa. The entries of $H_{i_1,i_2}$ correspond to edges cut by the bisection,
and minimizing the number of entries in $H_{i_1,i_2}$ corresponds to minimizing the
number of edges cut by the bisection (\emph{min-cut criterion}). This criterion is
often used in reordering matrices for parallel processing, where the 
off-diagonal matrix size determines the amount of communication between processors.

However, the number of entries in $H_{i_1,i_2}$ is not directly related to the
size of the matrix, as has been noted in the graph partitioning problem
for parallel computing \citep{Hendrickson1998a}. Instead, the size of the off-diagonal
matrix is given by the number of \emph{surface} vertices, i.e.~the number of vertices
that have cut edges. For this, we define a \emph{net} of a vertex $v$ in a graph 
$\mathcal{G}=(\mathcal{V},\mathcal{E})$  as \citep{Coon1995,Camarda1999}
\begin{equation}\label{eq:def_nets}
\text{net}(v)=\{u \in \mathcal{V} | u \text{ is adjacent to } v\}\,.
\end{equation}
Note that $v \in \text{net}(v)$, as $v$ is adjacent to itself. A net is said to
be cut by a bisection, if any two vertices $v_1,v_2\in \text{net}(v)$ are contained in different
sets $\mathcal{V}_{i_1}$ and $\mathcal{V}_{i_2}$. Then, the number of surface vertices and thus the size
of the off-diagonal matrix $H_{i_1,i_2}$ is given by the number of cut nets. Thus, minimizing
the number of cut nets (\emph{min-net-cut criterion}) corresponds to minimizing the 
the number of surface vertices, and thus to minimizing 
the size of the off-diagonal matrix $H_{i_1,i_2}$. 
Furthermore, since the vertices in $\mathcal{V}_{i_{1/2},\text{BFS}}$ are determined by a 
BFS emanating from the surface vertices, minimizing the number of cut nets will usually
also lead to a smaller number of vertices in $\mathcal{V}_{i_{1/2},\text{BFS}}$, leaving more
freedom towards achieving a balanced bisection. Figs.~\ref{fig:bisection_explanation}(b)
and (c) show a comparison of the min-cut and min-net-cut criterion for simple examples.
In practice, when minimizing the number of cut nets, we also use the min-cut
criterion to break ties between different bisections with the same number of
cut nets (\emph{min-net-cut-min-cut criterion}) in order to avoid wide local minima,
that occur frequently in the min-net-cut problem. 

Both the min-cut and min-net-cut bisection problem have been shown to be 
$\mathcal{NP}$-hard \citep{Garey1990}. Therefore, only heuristics are available
to solve them. These heuristics start from an initial (balanced) bisection, such
as constructed by the steps outlined above, and improve upon this initial bisection.
Here, we choose to use the Fiduccia-Mattheyses (FM) algorithm
\citep{Fiduccia1982}, as it is readily available for min-cut and min-net-cut bisection.
In fact, min-net-cut bisection is a \emph{hypergraph} partitioning problem, and 
the FM algorithm was originally designed for hypergraph partitioning. Furthermore,
the FM algorithm can naturally deal with locked vertices that may not be moved
between sets, is reasonable fast and its underlying concepts are easy to understand.
The FM heuristic is a \emph{pass-based} technique, i.e.~it is applied
repeatedly to the problem (several \emph{passes} are performed),
iteratively improving the bisection. More detailed information about the 
fundamentals of the Fiduccia-Mattheyses algorithm are given in
appendix \ref{sec:FM_explanation}.

We now summarize the steps outlined above and formulate an algorithm for
bisection:
\begin{alg}\label{bisection_algo}
Bisection of set $\mathcal{V}_i$ containing $N_i$ levels, with left (right) neighboring
set $\mathcal{V}_{i_\text{left}}$ ($\mathcal{V}_{i_\text{right}}$).
\begin{enumerate}
\item[A] Stop, if $N_i=1$.
\item[B] Do a BFS starting from $\mathcal{V}_{i_\text{left}}$ up to level 
$N_{i_1}=\text{Int}(N_i/2)$ and a BFS starting from $\mathcal{V}_{i_\text{right}}$ up to level 
$N_{i_2}=N_-\text{Int}(N_i/2)$. The vertices found by the BFS are assigned to $\mathcal{V}_{i_1}$
and $\mathcal{V}_{i_2}$, respectively, and are marked as locked.
\item[C] Distribute the remaining unassigned vertices taking into account the balance 
criterion \eqref{eq:balance_criterion}.
The vertices may be assigned according to either one of the following prescriptions:
\begin{enumerate}
\item[a)] Continue the BFSs from step B and
assign vertices to $\mathcal{V}_{i_1}$, if they are first reached by the BFS from 
$\mathcal{V}_{i_\text{left}}$, and to $\mathcal{V}_{i_2}$, if they are first reached by the BFS from 
$\mathcal{V}_{i_\text{right}}$. If a set has reached the size 
given by the balance criterion, assign all remaining vertices to the other set.
\item[b)] Distribute the unassigned vertices randomly to $\mathcal{V}_{i_1}$ and $\mathcal{V}_{i_2}$.
If a set has reached the size given by the balance criterion, 
assign all remaining vertices to the other set.
\end{enumerate}
\item[D] Optimize the sets $\mathcal{V}_{i_1}$ and $\mathcal{V}_{i_2}$ by changing the assignment
of unlocked vertices according to some minimization criterion. In particular,
the following optimizations may be performed:
\begin{enumerate}
\item[a)] No optimization. 
\item[b)] Min-cut optimization using the FM algorithm.
\item[c)] Min-net-cut optimization using the FM algorithm.
\item[d)] Min-net-cut-min-cut optimization using the FM algorithm.
\end{enumerate}
\end{enumerate}
\end{alg}

Recursive application of the bisection algorithm \ref{bisection_algo}
then leads to an algorithm for constructing a level set
complying with the requirements of the graph partitioning problem
\ref{graph_part_problem}, and thus an algorithm for block-tridiagonalizing
a matrix.
\begin{alg}\label{graph_part_algo}
Block-tridiagonalization of matrix $H$
\begin{enumerate}
\item[A] Construct the graph $\mathcal{G}=(\mathcal{V},\mathcal{E})$ corresponding to the matrix $H$,
and the sets $\mathcal{V}_0$ and $\mathcal{V}_{N+1}$ corresponding to the leads.
\item[B] Use algorithm \ref{BFSalgo} to determine the maximum number of
levels $N+2$. If $N<1$, stop.
\item[C] Construct $\mathcal{V}_1=\mathcal{V}\, \backslash\, (\mathcal{V}_0 \cup \mathcal{V}_{N+1})$, containing
$N$ levels.
\item[D] Apply the bisection algorithm \ref{bisection_algo} to
$\mathcal{V}_1$ and then recursively on the resulting subsets. Do not
further apply if a set only contains one level.
\end{enumerate}
\end{alg}
It should be emphasized, that the block-tridiagonalization does not require
any other input than the graph structure. In principle, the number of FM passes
may affect the result. However, from experience, this number can be
chosen as a fixed value, e.g.~10 FM passes, for all situations
\citep{Fiduccia1982}. Thus, the
block-tridiagonalization algorithm can serve as a black box. 

\begin{figure}
\begin{center}
\includegraphics[width=0.8\linewidth]{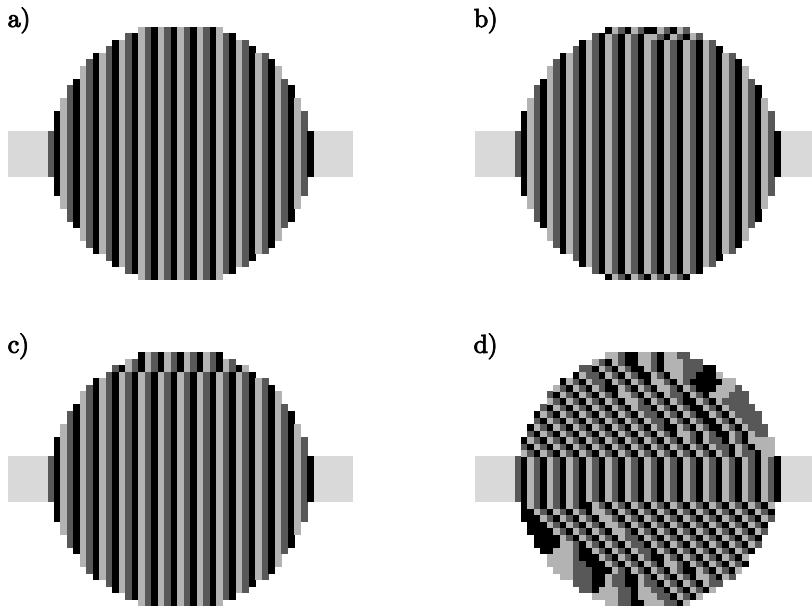}
\end{center}
\caption{Examples of level sets arising from (a)
the natural ordering of grid points (as in 
Fig.~\ref{fig:grids}), and application of the 
block-tridiagonalization algorithm \ref{graph_part_algo}
with distribution of vertices by BFS (algorithm \ref{bisection_algo},
step C.(a)) (b) without further optimization,
(c) with min-cut optimization, (d) with min-net-cut optimization.}
\label{fig:algo_application}
\end{figure}

In Fig.~\ref{fig:algo_application} we show for examples of level sets 
arising from the natural ordering of grid points (Fig.~\ref{fig:algo_application}(a),
\emph{natural level set})
and from the block-tridiagonalization algorithm developed
in this work (Fig.~\ref{fig:algo_application}(b)--(d)) for the case of a disk-type geometry. 
The level set in Fig.~\ref{fig:algo_application}(b) arises from recursive bisection, where the 
vertices were distributed according to a BFS without any optimization. The resulting
level set strongly resembles the natural level set. This is due to the highly symmetric
structure and the fact that vertices are assigned to levels
according to their distance from the leads---only small deviations are present 
due to the balance criterion. When the bisection is optimized according to the
min-cut criterion, Fig.~\ref{fig:algo_application}(c), the resulting level set changes 
little, as the min-cut criterion favors horizontal and vertical cuts 
for a square lattice, as presented in the example. In contrast, min-net-cut
optimization (Fig.~\ref{fig:algo_application}(d)) yields a new, non-trivial level set
that has less symmetry than the underlying structure. Note that the 
minimization of surface vertices leads to levels in the form
of ``droplets'', analogous to surface tension in liquids. 

In fact, we will show in 
Sec.~\ref{section:examples} that min-net-cut optimization usually
leads to level sets and thus block-tridiagonal orderings that 
are superior to those arising from other methods. In particular, they 
are better than the natural level sets, leading to a
significant speed-up of transport algorithms, as
demonstrated in Sec.~\ref{sec:examples_mesoscopics}. In addition to that,
the reordering algorithms allow one to use conventional two-terminal
transport algorithms also for more complicated, even multi-terminal
structures (see Secs.~\ref{sec:examples_mesoscopics} and
\ref{sec:examples_multiterminal}).

\subsubsection{Computational complexity}\label{sec:complexity_analysis}
 
We conclude the theoretical considerations with an analysis of the 
computational complexity of algorithms \ref{bisection_algo} and
\ref{graph_part_algo}. 

The bisection algorithm involves a BFS search on $\mathcal{V}_i$, which scales
linearly with the number of edges within $\mathcal{V}_i$, and thus has
complexity $\mathcal{O}(\abs{\mathcal{E}_i})$, where $\mathcal{E}_i$ is the set of edges within
$\mathcal{V}_i$. In addition to that, a single optimization pass of
the FM algorithm scales also as $\mathcal{O}(\abs{\mathcal{E}_i})$ \citep{Fiduccia1982}.
Usually, a constant number of passes independent of the size of the 
graph is enough to obtain converged results, and therefore the
optimization process of several FM passes is also considered
to scale as $\mathcal{O}(\abs{\mathcal{E}_i})$. Thus, the full bisection algorithm
also has complexity $\mathcal{O}(\abs{\mathcal{E}_i})$.

Usually, the number of edges per vertex is approximately
homogeneous throughout the graph. Since the  recursive bisection
is a divide-and-conquer approach, the computational
complexity of the full block-tridiagonalization algorithm is then
$\mathcal{O}(\abs{\mathcal{E}} \log\abs{\mathcal{E}})$ \citep{Sedgewick1992}. 
In typical graphs arising from physics problems, the number of edges
per vertex is a constant, the computational complexity can also
be written as $\mathcal{O}(N_\text{grid} \log N_\text{grid})$, where $N_\text{grid}$ is the number of
vertices in $V$, or the size of the matrix $H$.
 
In contrast, many quantum transport algorithms, such as the recursive
Green's function technique, scale as $\mathcal{O}(N (N_\text{grid}/N)^3)=\mathcal{O}(N_\text{grid}^3/N^2)$
in the optimal case of $N$ equally sized matrix blocks (levels) of size $N_\text{grid}/N$. Often,
the number of blocks (levels) $N\propto N_\text{grid}^{\alpha}$. Typically, to name a few examples,
 $\alpha=1$ in one-dimensional chains, $\alpha=1/2$ in two dimensions, and the
transport calculation scales as $\mathcal{O}(N_\text{grid}^{3-2\alpha})$. Thus, except for
the case of a linear chain, where $N=N_\text{grid}$ and matrix reordering is pointless anyways,
the block-tridiagonalization algorithm always scales more favorably than the 
quantum transport algorithms. This scaling implies that the overhead of the matrix
reordering in the transport calculation will become more negligible, 
the larger the system size. 

\section{Examples: Charge transport in two-dimensional systems}\label{section:examples}

\subsection{Ballistic transport in two-terminal devices}\label{sec:examples_mesoscopics}

We now evaluate the performance of the block-tridiagonalization algorithm
using representative examples from mesoscopic physics. The
Schr\"odinger equation for the two-dimensional electron gas (2DEG)
is usually transformed into a tight-binding
problem by the method of finite differences \citep{Kimball1934,Pauling1935,Frustaglia2004},
where the continuous differential equation is replaced by a set of linear equations
involving only the values of the wave function on discrete grid points. Commonly,
these points are arranged in a regular, square grid. This grid, together with the 
shape of the particular structure under consideration then defines the structure
of the Hamilton matrix and the corresponding graph.

\begin{figure}
\begin{center}
\includegraphics[width=0.8\linewidth]{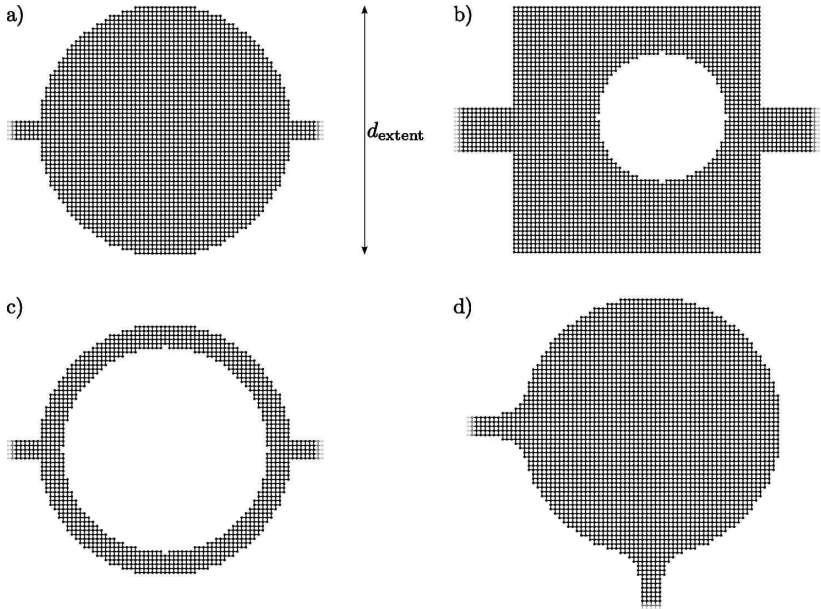}
\end{center}
\caption{Typical examples of structures considered in two-dimensional mesoscopic systems:
(a) circle billiard, (b) asymmetric Sinai billiard, (c) ring, and (d) circular cavity 
with perpendicular leads. The tight-binding grid arises from the finite difference
approximation to the Schr\"odinger equation. Note that the number of grid points
used here was deliberately chosen very small for visualization purposes.
In a real calculation, the number of grid points would be at least
2 orders of magnitude larger. $d_\text{extent}$ denotes a length characterizing the
extent for the different structures.}\label{fig:examples_2deg}
\end{figure}

The representative examples considered here are shown in Fig.~\ref{fig:examples_2deg}:
The circle (Fig.~\ref{fig:examples_2deg}(a)) and the asymmetric Sinai billiard 
(Fig.~\ref{fig:examples_2deg}(b)) that are examples of integrable and chaotic billiards 
in quantum chaos, the ring (Fig.~\ref{fig:examples_2deg}(c)) that may exhibit
various interference physics, and the circular cavity with leads that are
not parallel (Fig.~\ref{fig:examples_2deg}(d)) as an example of a structure
that does not have an intuitive, natural block-tridiagonal ordering. For all these
structures, we introduce a length scale $d_\mathrm{extent}$, given by the outer radius
of the circular structures and the side length of the square structure, characterizing
the maximum extent. The fineness of the grid, and thus the size of the corresponding
graph will be measured in number of grid points per length $d_\mathrm{extent}$.

We now apply the block-tridiagonalization algorithm using the various optimization criteria discussed
in the previous section, and compare the resulting orderings with the natural ordering and the 
ordering generated by the GPS algorithm. The weights $w(H)$, Eq.~\eqref{eq:matrix_weights},
of the different orderings are given in Table \ref{tab:weights}. 

\begin{table}
\caption{Weights $w(H)$, Eq.~\eqref{eq:matrix_weights}, 
for the block-tridiagonal ordering constructed by 
different algorithms for the examples of Fig.~\ref{fig:examples_2deg} . Optimization
was done by 10 passes of the FM algorithm, when the initial bisection
was constructed by BFS (algorithm \ref{bisection_algo}, step C.(a)), and 20 
passes, when the initial bisection was constructed by a random distribution 
of vertices (algorithm \ref{bisection_algo}, step C.(b)). 
The minimal weights for each system are printed bold. In all examples, there were 400 grid points per
length $d_\text{extent}$.}\label{tab:weights}
\begin{center}
\vspace{1em}
\begin{tabular}{>{\centering}m{0.22\linewidth}||>{\centering}m{0.16\linewidth}|>{\centering}m{0.16\linewidth}|>{\centering}m{0.16\linewidth}|>{\centering\arraybackslash}m{0.16\linewidth}}
&Circular billiard&Asymmetric Sinai billiard&Ring&Cavity with perp. leads\\
\hline\hline
natural block-\\tridiagonal ordering &$1.51\times10^{10}$&$1.58\times 10^{10}$&$8.72\times 10^{8}$&$-$\\
Gibbs-Poole-Stockmeyer&$1.15\times 10^{12}$&$7.84\times 10^{11}$&$2.14\times 10^{8}$&$7.05\times 10^{12}$\\
distribution by BFS, no optimization&$1.51\times 10^{10}$&$9.29\times 10^{9}$&$2.1\times 10^{8}$&$1.69\times 10^{10}$\\
distribution by BFS, min-cut&$1.51\times 10^{10}$&$9.67\times 10^{9}$&$2.1\times 10^{8}$&$1.59\times 10^{10}$\\
random distribution, min-cut&$2.22\times 10^{10}$&$9.95\times 10^{9}$&$2.1\times 10^{8}$&$5.13\times 10^{10}$\\
distribution by BFS, min-net-cut&$1.51\times 10^{10}$&$9.46\times 10^{9}$&$2.1\times 10^{8}$&$\mathbf{1.18\times 10^{10}}$\\
random distribution, min-net-cut&$1.46\times 10^{10}$&$\mathbf{9.0\times 10^{9}}$&$2.09\times 10^{8}$&$\mathbf{1.18\times 10^{10}}$\\
distribution by BFS, min-net-cut-min-cut&$\mathbf{1.26\times 10^{10}}$&$9.28\times 10^{9}$&$\mathbf{2.08\times 10^{8}}$&$1.24\times 10^{10}$\\
random distribution, min-net-cut-min-cut&$1.27\times 10^{10}$&$9.16\times 10^{9}$&$2.09\times 10^{8}$&$2.02\times 10^{10}$\\
\end{tabular}
\end{center}
\end{table}

The initial distributions for the bisection algorithm are done in two different ways: The vertices are
distributed both in an ordered way---by BFS---and randomly. The outcome after the optimization
however is always similar for both types of initial distributions which indicates
that the resulting weights are close to the global minimum and not stuck in a local minimum. Note that
we use twice as many FM passes for a random initial distribution than for an initial distribution 
by BFS, as convergence is usually slower for a random initial distribution.
 
In all examples, the min-net-cut criterion yields orderings with the best weights, as expected
from the considerations of the previous section. Based on the weight, orderings
according to this criterion are expected to give the best performance in transport calculations such as
the RGF algorithm. Note that the min-net-cut-min-cut ordering is on average closest to the
best ordering. The min-net-cut ordering sometimes suffers from slow convergence, when 
the algorithm must traverse a wide local minimum. The additional min-cut criterion helps 
to break ties and thus avoids these wide local minima. 

Except for the ring, where all algorithms perform well, the GPS algorithm 
yields weights that are even larger than the weight of the natural ordering.
As discussed above, the GPS algorithms performs well, if both leads are furthest apart in terms
of the graph. In the case of the ring, this is approximately fulfilled. 
In the general case, when the leads are at arbitrary positions, the GPS algorithm usually 
produces some very large levels. As the level size
enters cubically in the $w(H)$, this results in a prohibitively large weight. The GPS
algorithm thus cannot be used as a generic reordering algorithm for 
quantum transport according to problem \ref{graph_part_problem}.

In summary, the block-tridiagonalization algorithm \ref{graph_part_algo}
in the combination of initial distribution by BFS and min-net-cut-min-cut optimization 
yields the best reorderings with respect to the weight $w(H)$. Experience shows
that usually 10 FM passes are enough for optimizing a bisection. As a consequence, we
will use this combination exclusively in the rest of this work.

\begin{figure}
\begin{center}
\includegraphics[width=0.9\linewidth]{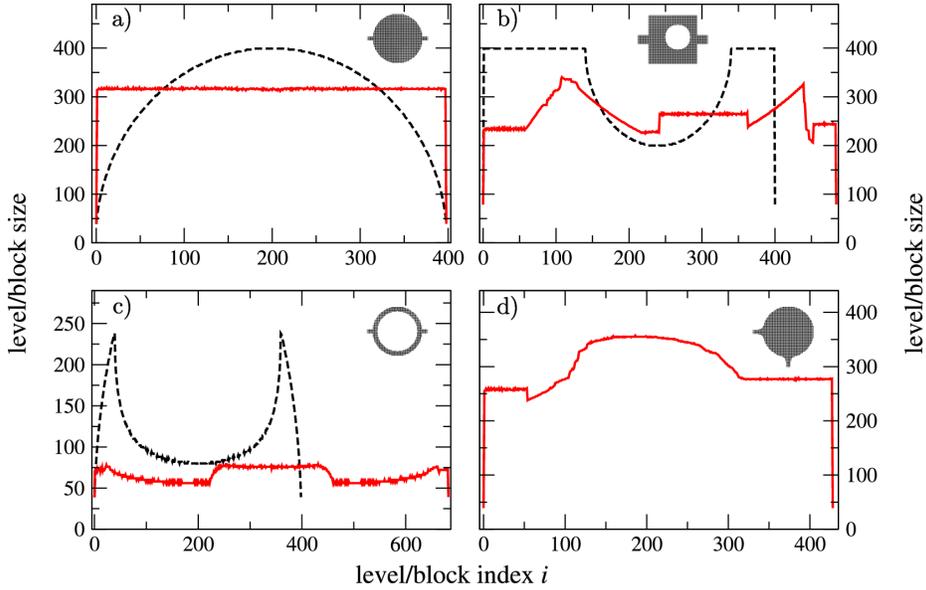}
\end{center}
\caption{Level (matrix block) size $M_i$ as a function of the level (matrix block) index $i$ for
the natural level set (dashed line) and the min-net-cut-min-cut reordering (solid line),
shown for (a) the circle billiard, (b) the asymmetric Sinai billiard, (c) the ring,
and (d) the circular cavity with perpendicular leads. Note that for (d), there is no
natural ordering. In all examples, there were 400 grid points per
length $d_\text{extent}$.}\label{fig:blocksizes}
\end{figure}

The weight $w(H)$ of a matrix is a global measure of the quality of a ordering. Additional
insight can be gained from the distribution of the sizes $M_i$ of the matrix blocks/levels.
In Fig.~\ref{fig:blocksizes} we show this distribution before and after reordering. For the
natural ordering of the finite difference grids, the number of matrix blocks is 
determined by the number of lattice points along the $x$-coordinate direction 
(see Fig.~\ref{fig:grids}(b)). In contrast, the number of matrix blocks after reordering
is given by the length of the shortest path between the two leads, in terms
of the corresponding graph.

In the case of the circle billiard, Fig.~\ref{fig:blocksizes}(a), the number of matrix blocks is the
same for the natural ordering and the reordered matrix, as the shortest path between the
leads is simply a straight line along the $x$-coordinate direction. The improvements
in the weight originate only from balancing the matrix block sizes: While the matrix block
sizes vary for the natural ordering---the lateral size changes along the
$x$-direction---the reordered matrix has equally sized matrix blocks. For this
particular example, the result of the block-tridiagonalization algorithm is optimal,
as it yields the best solution with respect to the requirements set forth in problems
\ref{matrix_reordering_problem} and \ref{graph_part_problem}. Note that in general
it is not always possible to find a perfectly balanced partitioning, but the circle
billiard is such an example.

In contrast, in the case of the asymmetric Sinai billiard and the ring the number of 
matrix blocks generated by the block-tridiagonalization algorithm is larger than in
the natural ordering (see Figs.~\ref{fig:blocksizes}(b) and (c), respectively). 
In both cases, the obstacle within the scattering region increases
the length of the shortest path connecting the two leads. In both examples,
this increase in the number of matrix blocks leads to a significantly decreased
weight $w(H)$ with respect to the natural ordering, although the partitioning is 
only approximately balanced. For instance, in the
particular case of the ring, the number of matrix blocks after reordering is approximately
given by the number of lattice points around half of the circumference. The
reordered ring thus has a weight very similar to a straight wire with a width
twice as large as the width of one arm of the ring, and a length given
by half of the ring circumference.

For the cavity with perpendicular leads, there is no natural ordering, and a 
specialized transport algorithm would be required. The 
reordering creates a matrix with approximately balanced block sizes, and
allows the direct application of conventional algorithms.

The weight $w(H)$ was introduced as a theoretical concept in order to
simulate the computational complexity of a transport calculation. After discussing the
influence of the reordering on this theoretical concept, we now demonstrate
how the reordering increases the performance of an actual quantum
transport calculation.

To this end we use a straight-forward implementation of the well-established
recursive Green's function algorithm for two terminals, as
described in Ref.~\citep{MacKinnon1985}. The necessary linear algebra 
operations are performed using the ATLAS implementation of LAPACK and BLAS
\citep{Whaley2005,Whaley2001}, optimized for specific processors. It should be emphasized 
that the code that does the actual transport calculation---such as calculation
of the Green's function and evaluation of the Fisher-Lee relation---is the same for all examples considered
here, including the non-trivial cavity with perpendicular leads. The abstraction
necessary for the reordering, i.e.~the graph structure and the corresponding
level set, allows for a generic computational code applicable to
any tight-binding model.

\begin{figure}
\begin{center}
\includegraphics[width=0.9\linewidth]{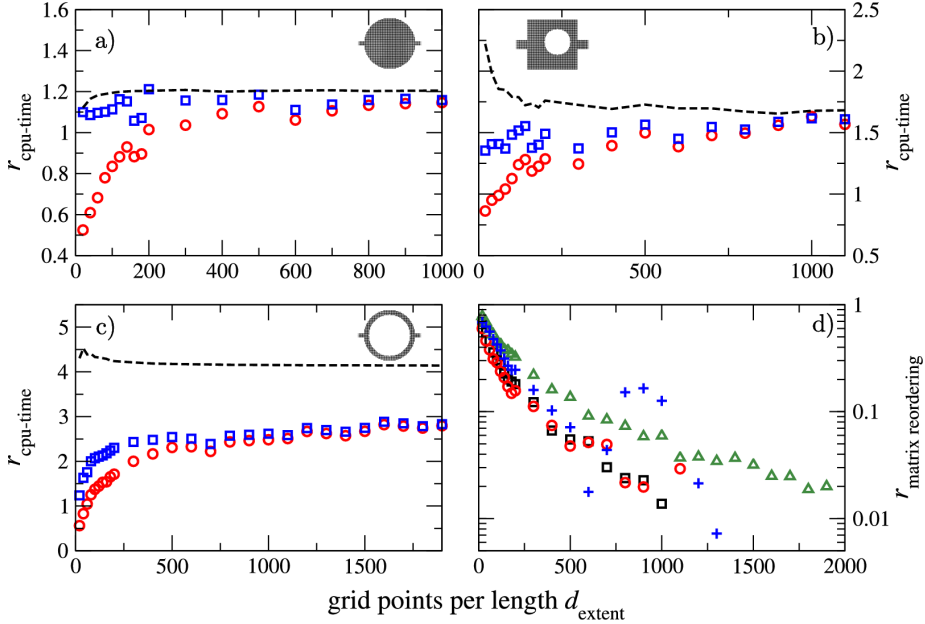}
\end{center}
\caption{(a)--(c): relative gain in computational time $r_\text{cpu-time}$,
Eq.~\eqref{eq:cpu_time}, through the
reordering as a function of the grid size for the circular billiard, the asymmetric Sinai billiard,
and the ring, respectively. $r_\text{cpu-time}$ is shown
excluding ($\square$) and including ($\bigcirc$) the overhead of matrix reordering.
The estimate for $r_\text{cpu-time}$ from the weights $w(H)$ of the different
orderings is shown as a dashed line. (d): fraction of time
$r_\text{matrix reordering}$, Eq.~\eqref{eq:matrix_reordering}, 
used for reordering the matrix as 
a function of the grid size. Data is shown for
the circular billiard ($\square$), the asymmetric Sinai
billiard ($\bigcirc$), the ring ($\triangle$), and the
circular cavity with perpendicular leads ($+$). The benchmarks were run on 
Pentium 4 processor with 2.8 GHz and 2 GBs of memory.}\label{fig:reordering_speedup}
\end{figure}

We measure the performance gain through matrix reordering as
\begin{equation}\label{eq:cpu_time}
r_\text{cpu-time}=\frac{\text{computing time for natural ordering}}
{\text{computing time for reordered matrix}}\,.
\end{equation}
Note that during a real calculation, the conductance
is usually  not only calculated once, but repeatedly as a function of some parameters,
such as Fermi energy or magnetic field. Thus, the respective quantum 
transport algorithm is executed repeatedly, too. In contrast, the block-tridiagonalization
has to be carried out again \emph{only} when the structure of the matrix
and thus the corresponding graph changes. For the examples considered here
this would correspond to changing the grid spacing or the shape of the structure.
In such a case, the overhead of matrix reordering must be taken into account
for $r_\text{cpu-time}$. This overhead can be quantified as
\begin{equation}\label{eq:matrix_reordering}
r_\text{matrix reordering}=\frac{\text{overhead of matrix reordering}}{\text{computing time including reordering}}\,.
\end{equation}
In a typical calculation however, the matrix structure given by the underlying 
tight-binding grid does not change, and the matrix reordering must be carried out
only once. In this common situation, the overhead of matrix reordering is negligible.
For example, any change of physical parameters such as Fermi energy, magnetic field
or disorder averages does not change the matrix structure.

In Fig.~\ref{fig:reordering_speedup} we show the performance gain through matrix
reordering, $r_\text{cpu-time}$, as a function of grid size for the circle billiard, the
asymmetric Sinai billiard, and the ring (Figs.~\ref{fig:reordering_speedup}(a)--(c), 
respectively). We include both measurements excluding and including the 
overhead of matrix reordering, as discussed above.
Remember that in the case of the cavity with perpendicular leads, 
Fig.~\ref{fig:examples_2deg}(d), there is no natural ordering and 
thus a performance comparison is not possible. 
In fact for this system, only matrix reordering makes a transport calculation 
possible in the first place. 

We find that block-tridiagonalization always increases the algorithmic
performance in the typical situation, when the overhead 
of matrix reordering can be neglected. However,
even if the reordering overhead is taken into account, we see a 
significant performance gain except for small systems---but there
the total computing time is very short anyway. In fact, as the system sizes 
increases, the overhead of reordering becomes negligible, as predicted
from the analysis of the computational complexity, and the performance
gains including and excluding the reordering overhead converge. This 
can also be seen in Fig.~\ref{fig:reordering_speedup}(d), where we show the 
reordering overhead $r_\text{matrix reordering}$ as a function of system size.

Especially for large systems, the total computing time can become very long, and any
performance gain is beneficial. Reordering leads to significant performance gains up to a 
factor of 3 in the case of the ring. The performance gain $r_\text{cpu-time}$
can also be estimated from the weights $w(H)$ of the original matrix (the natural
ordering) and the reordered matrix, shown as the dashed line 
in Figs.~\ref{fig:reordering_speedup}(a)--(c).
The actual, measured performance gain approaches this theoretical value, as
the system size increases. Note that we do not fully reach the theoretically predicted
performance gain in the case of the ring. On modern computer architectures,
computing time does not only depend on the number of arithmetic operations \citep{Whaley2001},
and thus the weight $w(H)$ overestimates the performance gain, though the
performance still improves significantly.

\begin{figure}
\begin{center}
\includegraphics[width=0.5\linewidth]{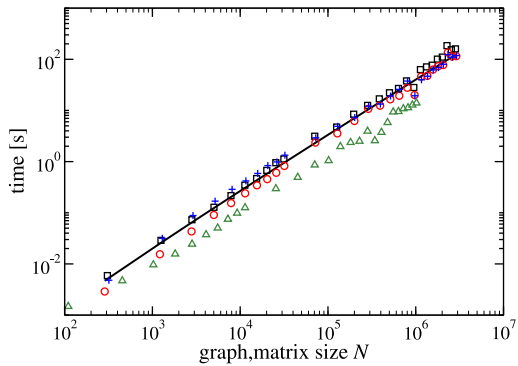}
\end{center}
\caption{Time spent for matrix reordering as a function of the total grid
(matrix) size $N$, for the circular billiard ($\square$), the asymmetric Sinai
billiard ($\bigcirc$), the ring ($\triangle$), and the
circular cavity with perpendicular leads ($+$). The solid line
is a fit to the predicted scaling of the computational complexity,
$N \log N$.}\label{fig:reordering_scaling}
\end{figure}

Finally, we demonstrate the $\mathcal{O}(N_\text{grid} \log N_\text{grid})$ 
scaling of the reordering algorithm.
Fig.~\ref{fig:reordering_scaling} shows the computing times of the 
block-tridiagonalization algorithm as a function of matrix/graph size $N$ for
the geometries considered in this section. For all systems, the computing times
scale according to the prediction from the complexity analysis in 
Sec.~\ref{sec:complexity_analysis}, as apparent from the fit
$\propto N_\text{grid} \log N_\text{grid}$. Note that for large $N_\text{grid}$, 
$\mathcal{O}(N_\text{grid} \log N_\text{grid})$ scaling is practically 
indistinguishable from $\mathcal{O}(N_\text{grid})$,
as can also be seen in Fig.~\ref{fig:reordering_scaling}.

In the examples of this section, we considered the pedagogic
case of charge transport on a square, finite difference grid. The approach
presented here can however immediately applied to more complex situations,
such as spin transport, as reviewed in Ref.~\citep{Wimmer2008a}. 
In addition, extending the transport calculation to a different grid
is straightforward, as any tight-binding grid can be encoded into a graph,
and the block-tridiagonalization algorithm has already been applied to the case of
the hexagonal grid of graphene \citep{Wimmer2008} 
(for a review on graphene see \citep{CastroNeto2008}). 
A further example of this versatility is shown in the
next section, where we apply the block-tridiagonalization algorithm
to solve multi-terminal structures involving different tight-binding models.

\subsection{Multi-terminal structures}\label{sec:examples_multiterminal}

In the previous section, we demonstrated that matrix reordering increases the performance
of quantum transport algorithms for two-terminal structures and additionally
makes it possible to apply these conventional algorithms to non-trivial structures. Whereas
there is a great variety of quantum transport algorithms for systems with two leads, 
there are only few algorithms that are suitable for multi-terminal structures, and
most of these are restricted to rather specific geometries (e.g.~Ref.~\citep{Baranger1991}).
Only recently algorithms have been developed that claim to be applicable to any multi-terminal
structure. The \emph{knitting algorithm} of Ref.~\citep{Kazymyrenko2008} is a variant of the RGF 
algorithm where the system is built up adding every lattice point individually, instead
of adding whole blocks of lattice points at a time. Therefore, instead of a matrix multiplication, the
central computational step is an exterior product of vectors. Unfortunately, 
this implies that the knitting algorithm cannot use highly optimized
matrix multiplication routines (Level 3 BLAS operations), that are usually much more
efficient than their vector counterparts (Level 2 BLAS operations), as
discussed in Ref.~\citep{Whaley2001}. Another multi-terminal transport algorithm presented
recently \citep{Qiao2007}, is based on the transfer matrix approach. However, it requires
the Hamiltonian to be in a specific block-tridiagonal form, and the corresponding level set
is set up manually.

\begin{figure}
\begin{center}
\includegraphics[width=0.8\linewidth]{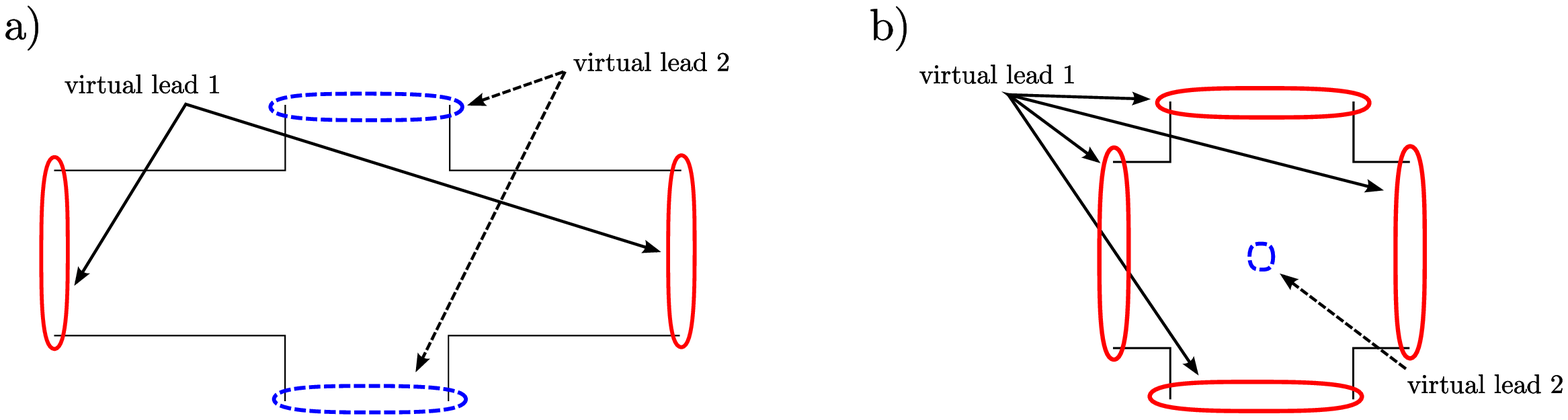}
\end{center}
\caption{A multi-terminal structure can be reduced to an equivalent
two-terminal structure by collecting all leads in two
\emph{virtual leads}. (a) The leads are redistributed into two virtual leads. (b) All leads are
combined in a single virtual lead, the second virtual lead is formed by a 
vertex furthest away.}\label{fig:multiterminal_virtualleads}
\end{figure}

Here we show how to employ the block-tridiagonalization algorithm in order to 
apply the well-established two-terminal quantum transport algorithms to 
an arbitrary multi-terminal system. The basic idea is sketched in 
Fig.~\ref{fig:multiterminal_virtualleads}(a): Combining several
(real) leads into only two virtual leads the multi-terminal problem is
reduced to an equivalent two-terminal problem. After reordering, the
resulting problem can then be solved by conventional two-terminal
algorithms. 
Note that in this approach the number of matrix blocks is given by the shortest
path between leads in two different virtual leads. If all leads are very close
together, this may lead to only few, large blocks in the reordered matrix and respectively
levels in the graph partitioning, leading to a very large weight $w(H)$. In such a
case it is advisable to collect all leads into a single virtual lead. The second
virtual lead is then formed by a vertex in the graph, that is furthest away from all
leads as depicted in Fig.~\ref{fig:multiterminal_virtualleads}(b). 
Such a vertex can be found by a BFS search originating from all leads. Thereby
the number of matrix blocks/levels is maximized. In fact, this approach
yields a block-tridiagonal matrix structure as required by the algorithm of Ref.~\citep{Qiao2007}.

\begin{figure}
\begin{center}
\includegraphics[width=0.7\linewidth]{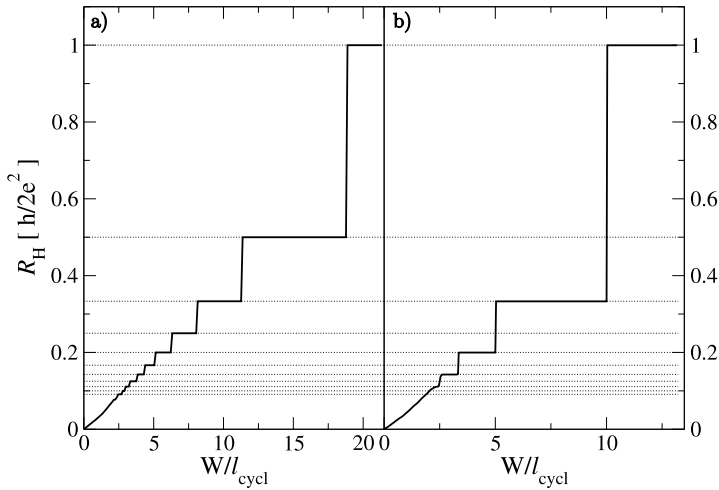}
\end{center}
\caption{Example of a four-terminal calculation: Quantum
hall effect (a) in a two-dimensional electron gas
and (b) in graphene. The Hall resistance $R_\text{H}$ is shown as 
a function of $W/l_\text{cycl}$, where $W$ is the width of the
Hall bar and $l_\text{cycl}$ the cyclotron radius in a magnetic field $B$.
Note that $W/l_\text{cycl}\propto B$. The dotted lines indicate the quantized
values of the Hall resistance, $h/2 e^2 \times n^{-1}$, where $n$ is 
a positive integer.}\label{fig:example_halleffect}
\end{figure}

We now demonstrate these strategies on the example of the quantum Hall effect (QHE) in
a 2DEG formed in a semiconductor heterostructure \citep{Klitzing1980}
and in graphene \citep{Novoselov2005}. For this we use a four-terminal Hall bar 
geometry as sketched in Fig.~\ref{fig:multiterminal_virtualleads}(a), on top
of a square lattice (finite difference approximation to 2DEG) and a hexagonal
lattice . Again, it should
be emphasized that the code of the actual transport calculation is the
same as employed in the two-terminal examples of the previous section. 
The results of the calculation are shown in Fig.~\ref{fig:example_halleffect},
where the integer QHE of the 2DEG and the odd-integer QHE of graphene are clearly
visible. 

The methods outlined above make it possible to calculate quantum transport
in \emph{any} system described by a tight-binding Hamiltonian. This generality
is one of the main advantages gained by using the matrix reordering. However, generality
also implies that it is difficult to make use of properties of specific systems,
such as symmetries, in order to speed up calculations. Special algorithms
developed specifically for a certain system however can, and will usually
be faster than a generic approach---at the cost of additional development time.

In the case of the Hall geometry in a 2DEG, such a special algorithm was
presented by Baranger et al.~\citep{Baranger1991}, and we have implemented
a variant of it. Comparing the computing times for the Hall bar geometry in
a 2DEG, we find that the special algorithm is only a factor of $1.6-1.7$ faster
than our generic approach. Although such a performance comparison
may depend crucially on the details of the system under consideration,
experience shows that the use of the generic approach often does not
come with a big performance penalty.

\section{Conclusions}\label{section:conclusions}

We have developed a block-tridiagonalization algorithm based on graph partitioning
techniques that can serve as a preconditioning step for a wide
class of quantum transport algorithms. The algorithm can be applied to any
Hamilton matrix originating from an arbitrary tight-binding formulation
and brings this matrix into a form that is more suitable for many two-terminal
quantum transport algorithms, such as the widely used 
recursive Green's function algorithm. The advantages
of this reordering are twofold: First, the reordering can speed up 
the transport calculation significantly. Second, it allows for applying conventional two-terminal
algorithms to non-trivial geometries including non-collinear leads
and multi-terminal systems. The block-tridiagonalization algorithm
scales as $\mathcal{O}(N_\text{grid} \log N_\text{grid})$, where 
$N_\text{grid}$ is the size of the Hamilton matrix, and thus
induces only little additional overhead. We have demonstrated the performance
of the matrix reordering on representative examples, including transport
in 2DEGs and graphene.

The block-tridiagonalization algorithm can operate as a black box and
serve as the foundation of a generic transport code that can be applied
to arbitrary tight-binding systems. Such a generic transport code
is desirable, as it minimizes development time and increases code quality, 
as only few basic transport routines are necessary, that can be tested thoroughly.

We acknowledge financial support from DFG within GRK638 and SFB689.

\appendix

\section{The Fiduccia-Mattheyses algorithm}\label{sec:FM_explanation}

\subsection{Graphs and hypergraphs}

The Fiduccia-Mattheyses algorithm was originally developed for hypergraph
partitioning \citep{Fiduccia1982}. A hypergraph $\mathcal{H}$ is an ordered pair 
$\mathcal{H}=(\mathcal{V}, \mathcal{N})$, where $\mathcal{V}$ is a set
of vertices, and $\mathcal{N}$ a set of \emph{nets} $n_i$ (also called 
\emph{hyperedges}) between them. A net $n_i$ is
a set of vertices, i.e.~$n_i \subset \mathcal{V}$. An undirected graph is
a special realization of a hypergraph, where every net contains
exactly two vertices. Thus, any algorithm for hypergraph
partitioning can also be applied to an undirected graph.

\begin{figure}
\begin{center}
\includegraphics[width=0.6\linewidth]{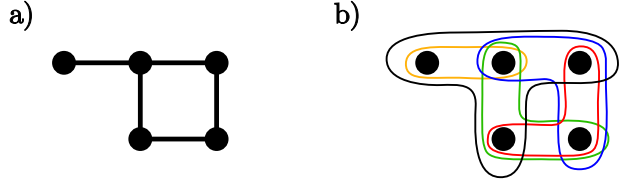}
\end{center}
\caption{Schematic representation of (a) a simple graph and (b) the 
corresponding hypergraph structure imposed through all nets, 
$\mathcal{N}=\{\text{net}(v)\, |\, v \in \mathcal{V}\}$.}\label{fig:example_graph_hypergraph}
\end{figure}

During the FM bisection, we have to consider the graph structure
arising from the Hamiltonian matrix in order to minimize the
number of cut edges (min-cut), whereas for minimizing the
number of surface vertices, i.e.~the number of cut nets (min-net-cut),
the hypergraph structure arising from all nets $\text{net}(v)$
as defined in Eq.~\eqref{eq:def_nets}, 
$\mathcal{N}=\{\text{net}(v)\, |\, v \in \mathcal{V}\}$, is essential.
For min-net-cut-min-cut optimization, we have to consider both structures simultaneously.
A schematic representation of a graph and the corresponding hypergraph
structure is shown in Fig.~\ref{fig:example_graph_hypergraph}.

\subsection{Fiduccia-Mattheyses bisection}

The FM algorithm is based on the concept of \emph{gain}. The gain of a vertex
in an existing bisection is defined as the change in weight, 
i.e.~the number of cut edges or nets, that occurs when this vertex 
is move to the other part. This gain can also be negative, if
such a move increases the number of cut edges or nets. The basic idea
of the FM algorithm is to swap vertices with the highest gain between parts, while
obeying some balance criterion. The fact that the highest gain can be negative,
helps the FM algorithm to escape local minima.
After moving, the respective vertex is locked
in order to avoid an infinite loop, where a single vertex might be swapped
back and forth repeatedly. The FM pass ends, when all (free) vertices have been moved,
and the best bisection encountered during the pass is returned as result. Further 
passes can then successively improve on this bisection.

\end{document}